\documentclass[ twocolumn,aps,prd,   
               preprintnumbers,numbers,sort&compress,nofootinbib,
                            showpacs,
               colorlinks,
               linkcolor=blue,
               citecolor=blue]{revtex4-1}
   \newcommand{\exclude}[1]{}

\usepackage{graphicx,amsmath,amssymb,bm}
\usepackage{psfrag}
\usepackage{feynmp}
\usepackage{hyperref}
\usepackage{enumitem}

 \usepackage{moreverb}
\usepackage{subfigure}
\usepackage{multirow}
\usepackage{braket}
\usepackage{amsfonts}
\usepackage{breqn}
\usepackage{float}

\newcommand{\beq}{\begin{equation}}
\newcommand{\eeq}{\end{equation}}
\newcommand{\be}{\begin{eqnarray}}
\newcommand{\ee}{\end{eqnarray}}
   
\def\dd{ \,\mathrm{d} }
\def\d{\partial}
\def\+{\dagger}
 \def\la{\langle}
 \def\ra{\rangle}

\begin{document}

\title{Topological Casimir effect in Maxwell Electrodynamics on a Compact Manifold}

\author{ChunJun Cao, Moos van Caspel \&  Ariel R. Zhitnitsky}

 \affiliation{Department of Physics \& Astronomy$,$   University of British Columbia$,$  Vancouver$,$    Canada.}
 
 \date{\today}

\begin{abstract}
 We study the Topological Casimir effect, in which extra  vacuum  energy emerges  as a result of the topological features of the theory, rather than  due to the conventional fluctuations 
 of the physical  propagating  degrees of freedom.   
We compute the corresponding topological   term in quantum Maxwell theory defined on a compact manifold.   Numerically, the topological effect  is  much smaller than the  conventional Casimir effect.
   However,  we argue that  the Topological Casimir  Effect is highly sensitive to  an external magnetic field, which may help to discriminate it from the conventional Casimir effect. It is quite amazing that the external magnetic field plays the role  of the  $\theta$ state, similar to a $\theta$ vacuum in QCD, or $\theta=\pi$ in topological insulators.

\end{abstract} 
\maketitle

\section{Introduction. Motivation.}\label{introduction}
The nature  of the conventional Casimir energy is well-understood by now:
the effect is due to the vacuum fluctuations of physical photons, which have slightly different 
propagating properties in the presence  of boundaries in comparison with infinite Minkowski space. Essentially, the electromagnetic modes get modified  as a result of nontrivial  boundary conditions (BC). 
For the well-known example of parallel conducting plates, this tiny deviation leads to the well known expression for the Casimir energy
\be
\label{energy}
E_{\rm C}\equiv \left(E_{\rm BC}-E_{\rm Minkowski}\right)=-\frac{L^2\pi^2}{720 a^3},
\ee
where $a$ is the separation distance between the two plates of size $L$. 
This extra energy gives rise to an attractive force per unit area (vacuum pressure) \cite{Casimir}
\be
\label{pressure}
P=-\frac{\d \left(E_{\rm C}/L^2\right)}{\d a}=-\frac{\pi^2}{240 a^4}.
\ee
Today the Casimir force has been measured \cite{experiment}, confirming Casimir's basic idea. 

Since its original prediction, the Casimir effect has been studied for countless configurations with fields of various spins. The Casimir effect on nontrivial topological spaces has also been widely explored. In many such cases, a simple scalar field is considered and periodic or twisted boundary conditions are imposed to reflect the topological properties. The fluctuations of the physical field are quantized, for instance, by the periodicity of the space, yielding a Casimir energy similar to (\ref{energy}). See \cite{review} for an overview.

However, in the case of gauge fields, we argue that it is important to not only account for the topology of the spacetime manifold, but also the relation between that and the gauge topology. Precisely this topology of the gauge group leads to the emergence of vacuum states that are physically identical but topologically inequivalent. These are known as winding states and are often overlooked in literature on the Casimir effect.

We will explicitly demonstrate in the present work that, for the Casimir effect formulated using a pure photon field on a spacetime manifold with toroidal topology, the non-trivial spatial and gauge topology together induce an additional vacuum pressure that has not been previously computed. Such an effect is purely topological in origin, resulting not from fluctuations of the physically propagating degrees of freedom as in the ``conventional'' Casimir effect, but rather from the tunnelling between different topological sectors. Mathematically, such phenomena are described by the fundamental group $\pi_1[U(1)]\cong \mathbb{Z}$, where non-trivial mappings between the spacetime manifold and the gauge group assume the form of gauge transformations.
Due to its topological nature, the extra contribution has some unique qualities, both theoretical and practical, that distinguish it from the conventional Casimir effect. 

  A simple way to get some feeling on the nature of these new topological contributions is to study Maxwell theory in two dimensions, which is essentially the Schwinger model without fermions. As is well known, Maxwell theory in two dimensions is empty, since there are no physical propagating degrees of freedom. Still, there are nontrivial topological sectors  in the model which eventually lead to the emergence  of the so-called $\theta$ vacuum state. The construction of these topological sectors is sensitive to the size of the system. Therefore, it is not a surprise that the partition function  will be also sensitive to 
  the system's size, including finite size along the compactified time direction $\beta$,  corresponding to a finite temperature $T=1/\beta$. We will elaborate on this example in great detail in section \ref{topology}, using the Hamiltonian as well as Euclidean path integral approaches to explain the nature  of topological vacuum fluctuations. We also elaborate on the physical ``reality" of these vacuum fluctuations in Appendix \ref{2d} where we compute some important observables, such as topological susceptibility and entropy. 

We also note here that the gauge-induced contributions to the Casimir energy on a compact manifold 
 have been discussed previously in literature. 
Although no explicit computations were performed, it has been suggested in \cite{70stheory} that a sum over all gauge classes may be required to accommodate the non-trivial topological features of the theory. 
More recently, the computations for the electromagnetic field on general manifolds were explicitly done in \cite{Kelnhofer:2012ig}, and papers referenced therein. Our goal here is to discuss some  key  elements of these new topological terms in a more physical and intuitive way, rather than through formal mathematics. Furthermore,   we will   discuss the relation   between   the  $\theta$ states  and the physical  realization of these states by placing  the system into a uniform external   magnetic field.   We speculate that a high  sensitivity of the extra terms to the applied external magnetic field might be a key element which could  allow one to measure these novel types of vacuum fluctuations in real experiments.

 To conclude this introduction, we wish to comment on the title of this work and the term   ``Topological Casimir Effect" (TCE) which will be frequently  used in the text below. In some literature this name can refer to the conventional Casimir effect on different topological spaces. However, in the current context, we will use it to strictly denote the additional topological contribution from tunnelling phenomena between the nontrivial topological sectors that make up the $\theta$-state, which is the true vacuum of the configuration. The effect is fundamentally different from that obtained by solely manipulating the spacetime topology and is unique to gauge fields. Exactly in this context, this term was introduced in \cite{Zhitnitsky:2011aa} to emphasize that new extra contribution to the vacuum energy may emerge as a result of tunnelling events. 
 
 
The structure of our presentation is as follows. In the next section, we review the relevant parts of the two dimensional Maxwell ``empty" theory which does not have  any physical propagating degrees of freedom, but does show nontrivial topological features.  We study  this system using the Euclidean path integral approach as well as the Hamiltonian formalism. In section \ref{4d} we generalize our construction to four dimensional Maxwell theory. Numerical estimates in this case suggest that TCE  is generally much smaller than the conventional CE in normal circumstances. However, in section \ref{B} we advocate an idea  that the effect  is highly sensitive to a weak uniform external magnetic field. It is very similar  to a construction of the    so-called $\theta$ states in QCD.     Finally, in section \ref{conclusion} which is our conclusion, we comment on some profound consequences   the Topological Casimir Effect  may have for cosmology. Furthermore, we advocate the idea that an experimental    study  of the 
Topological Casimir Effect  in a laboratory might  be considered as an  investigation of the most intricate properties of the cosmological vacuum and the dark energy observed in our universe. In Appendix \ref{2d} we argue, using an ``empty"  two dimensional Maxwell model, that the topological vacuum fluctuations are very real and very physical and must be taken into consideration to satisfy some important consistency conditions  such as the Ward Identities. 
  
  \section{Maxwell theory in two dimensions}\label{topology}
  The 2d Maxwell model has been solved numerous times using very different techniques, see e.g. \cite{Manton:1985jm,Balachandran:1994vi, SW} for a review. 
  We have nothing new to say here. Our goal is in fact quite different: we want to review this ``empty" model by emphasizing some elements which will be crucial for our discussions of the Topological Casimir Effect in   four dimensions. 
  \subsection{Hamiltonian  framework}\label{hamiltonian}
  We consider 2d Maxwell theory defined on the Euclidean torus $S^1\times S^1$ with lengths $L$ and $\beta$ respectively. In the Hamiltonian  framework we choose a $A_0=0$ gauge along with $\partial_1 A_1=0$. This implies that $A_1(t)$ is the only dynamical variable of the system with $E=\dot{A_1}$. The Hamiltonian density, the Gauss law and the commutation relations are
  \be
  \label{1}
 {\cal{H}}=\frac{1}{2}E^2, ~\partial_1 E |{\rm phys.} \rangle=0,\\ \nonumber
  [A_1(x), E(y)]=i\hbar\delta(x-y),
  \ee
  where $|{\rm phys.} \rangle$ is the physical subspace. The Gauss law is satisfied only for the $x$-independent (zero) mode.   Therefore, the problem is reduced to the quantum mechanical (QM) problem  of a single zero mode living on a circle of circumference $ L$. In other words, the configurations
     \be
     \label{2}
  A_1\approx A_1+ \frac{2\pi n}{eL}, ~~~~ n\in  \mathbb{Z}
  \ee
  are gauge equivalent and must be identified.  The fact that 2d Maxwell theory does not describe any physical propagating degrees of freedom is well known-- it simply follows from the observation  that the polarization of a photon   must be perpendicular to its momentum. However, such a polarization  can not live in the physical space as there is only one spatial dimension $x$, which is reserved for momentum. The presence of a single $x$-independent mode and the absence of all other $x$-dependent modes are manifestations of the ``emptiness" of this theory.  
  
  The loop integral $e\int dx A_1=eA_1L$ plays the role of  phase 
  $\phi$ in the conventional  QM problem for a particle on a circle with periodic boundary conditions. 
 The commutation relation (\ref{1}) then implies  that the
 electric field $E$ is a constant in space and that it is quantized:
 \be
 \label{E}
 E= en ~~~ n\in   \mathbb{Z}. 
 \ee
 The hamiltonian $ H\equiv {\cal{H}}L $ and the corresponding eigenvalues $E_n$ for this system are well known and are given by 
   \be
  \label{3}
   H=-\frac{1}{2L}\cdot\frac{d^2}{dA_1^2}, ~~~ E_n= \frac{1}{2}n^2e^2 L.
  \ee
  Consequently, the partition function for this system  is
 \be
 \label{Z_1}
 {\cal{Z}}(\beta, L)=\sum_{n\in \mathbb{Z}} e^{-\beta E_n}=\sum_{n\in \mathbb{Z}} e^{-\frac{1}{2}\beta L n^2e^2}.
 \ee
The  construction of  the so-called $\theta$ states  is also  well known  for this system \cite{Manton:1985jm}. 
  The spectrum in this case  is  shifted as follows $E_n(\theta)= \frac{1}{2}\left(n+\frac{\theta}{2\pi}\right)^2e^2 L$, such that 
  the corresponding  partition function now takes the form
  \be
 \label{Z_2}
 {\cal{Z}}(V, \theta)=   \sum_{n\in \mathbb{Z}} e^{-\frac{e^2V}{2} \left(n+\frac{\theta}{2\pi}\right)^2},
 \ee
 where $V = \beta L$ is the two-volume of the system.
  Before we discuss the physical meaning of the obtained results in the context of our present work, we want to reproduce the same  partition function for the same 2d Maxwell theory using the  path integral approach. In this case, the interpretation of eq. (\ref{Z_2}) will be quite obvious and straightforward. Furthermore, it can easily be generalized to four dimensional Maxwell theory defined on a compact manifold. 
  
  \subsection{Euclidean Path Integral Approach}\label{pathintegral}
For path integral computations, we use a Wick rotation to desribe the system in a Euclidean metric. Here the inverse temperature $\beta = 1/T$ takes the role of an imaginary time component with periodic BC, such that we can consider a two dimensional Euclidean torus $\beta \times L$. We follow \cite{SW} and   introduce the  classical ``instantons" in order to describe   the  different  topological sectors of the theory  which are classified  by the integer $k$.
The transitions between different topological $k$-sectors are described by these ``instantons", as given by the following configuration\cite{SW}:
 \be
 \label{inst}
   e E^{(k)}=\frac{2\pi k}{V}, 
 \ee
 where $Q=\frac{e}{2\pi}E$ is the topological charge density 
and 
\be
\label{Q}
\int  \dd^2x ~Q(x)= \frac{e}{2\pi} \int \dd^2x ~E(x) =k
\ee
is the integer-valued topological charge   in the 2d $U(1)$ gauge theory, $E(x)=\partial_0A_1-\partial_1A_0$ is the field strength\footnote{One should not confuse the electric field from the Euclidean formulation (\ref{inst}) with an $E$ field (\ref{E})
computed in Minkowski space. In the former case it is an unphysical complex configuration saturating the path integral, while in the latter it is a ``real" physical fluctuating electric field in the Hamiltonian formalism in Minkowski space.}. 
The action of this classical configuration is
 \be
 \label{action}
 \frac{1}{2}\int d^2x E^2= \frac{2\pi^2 k^2}{e^2 V}.
 \ee
   This configuration corresponds to the   topological charge $k$ as defined by (\ref{Q}).
The next step is to   compute the  partition function defined as follows
\be
\label{Z_3}
{\cal{Z}}(\theta)=\sum_{ k \in \mathbb{Z}}{\int {\cal{D}}}A^{(k)} {e^{-\frac{1}{2}\int d^2x E^2+i  \frac{e\theta }{2\pi}  \int d^2x E}}.
\ee
   All integrals in this partition function are gaussian and can be easily evaluated using the technique developed in \cite{SW}. The result is
    \be
 \label{Z_4}
 {\cal{Z}}(\beta, L, \theta)= \sqrt{\frac{2\pi}{e^2V}}\sum_{k\in \mathbb{Z}} e^{-\frac{2\pi^2k^2}{e^2V} +ik\theta},
    \ee
    where the expression in the exponent represents the classical instanton configurations with action (\ref{action}) and topological charge (\ref{Q}), while the factor in front is due to the fluctuations. 
The computation of this   pre-exponent  factor  is reduced to a conventional quantum mechanical (QM) problem
as the fluctuating field is in fact  $x$-independent in the $A_0=0$ gauge, as mentioned in section \ref{hamiltonian}.
Therefore, the expression for the pre-exponent  is  
\be
\label{Z_5}
 \int {\cal{D}} (\delta A_1) e^{-\frac{L}{2}\int^{\beta}_0 d\tau  ( \delta\dot{ A}_1)^2}  
 \ee
 A simple way to evaluate this path integral is to rescale the $A_1$ field according to its natural dimensionality $A_1\equiv a_1\left(\frac{2\pi}{eL}\right)$
 where the dimensionless variable $ 0\leq a_1\leq 1$ fluctuates inside a unit  interval according to (\ref{2}).
 In terms of this rescaled field problem, (\ref{Z_5}) is reduced to a standard expression for a free particle with mass $m\equiv L\left(\frac{2\pi}{eL}\right)^2$ such that
 \be
\label{Z_6}
 \int {\cal{D}} \delta a_1 e^{-\frac{L}{2}\left(\frac{2\pi}{eL}\right)^2\int^{\beta}_0 d\tau  ( \delta\dot{ a}_1)^2} =
 \sqrt{\frac{m}{2\pi\beta}}=   \sqrt{\frac{2\pi}{e^2L\beta}}~~~
 \ee
 which is precisely the  pre-exponent   factor  in formula (\ref{Z_4}).
   
   While expressions (\ref{Z_2}) and (\ref{Z_4}) look differently, they are actually identically the same, as the Poisson summation formula states:
     \be
 \label{poisson}
  \sum_{n\in \mathbb{Z}} e^{-\frac{e^2V}{2} \left(n+\frac{\theta}{2\pi}\right)^2}= \sqrt{\frac{2\pi}{e^2V}}\sum_{k\in \mathbb{Z}} e^{-\frac{2\pi^2k^2}{e^2V} +ik\theta},
    \ee
  see  \cite{Azakov:2005qn} with detailed discussions on the relation between Hamiltonian formalism and the path integral approach. 
   \subsection{Interpretation}\label{Interpretation}
  The crucial observation for our present study is that this naively ``empty" theory which has no physical propagating degrees of freedom, nevertheless shows some very nontrivial features of the ground state related to the topological properties of the   theory. These properties   are inherent features of the gauge theories and do not have counterparts  in conventional scalar field theories. Rather these new properties are related to the presence of different topological sectors 
in the system, which we refer to as the ``degeneracy" of the ground state, for short\footnote{\label{degeneracy}Not to be confused with the conventional term ``degeneracy", when two or more physically distinct states are present in the system. In the context of this paper the
     ``degeneracy" implies the existence of winding states $| n\ra$ constructed as follows: ${\cal T} | n\ra= | n+1\ra$.  In this formula the operator ${\cal T}$ is  the  large gauge transformation operator  which commutes  with the Hamiltonian $[{\cal T}, H]=0$. 
     The physical vacuum state is {\it unique}
     and constructed as a superposition of $ |n\ra$ states as follows $| \theta\ra= \sum \exp(in\theta) |n\ra$. In the path integral approach, the presence of $n$ different sectors in the system is reflected  by  summation over $ { k \in \mathbb{Z}}$ in eq. (\ref{Z_3},\ref{Z_4}). }.    We interpret the nontrivial properties of the partition function (\ref{Z_3},\ref{Z_4})  in this ``empty" model as a result of tunnelling between these ``degenerate" winding $ |n\ra$ states.  These tunnelling processes are happening all the time and the intensity of tunnelling is determined by the topological charge (\ref{Q}) and the size $V$ of the compact manifold (\ref{action}). A typical value of the topological charge $k$ which saturates the series (\ref{Z_4}) in the large volume limit is very large, $k\sim \sqrt{e^2 V}\gg 1$. 
     
     It is different from the conventional tunnelling in QM in that the tunnelling in our system corresponds to a transition between  
     one and the same physical state, whereas that in QM describes a transition between physically distinct states. More specifically, the tunnelling in our case occurs between the winding $ |n\ra$ states
     which are connected by large gauge transformations. Therefore, they correspond to one and the same physical state. 
     
     The key for our present work is the observation that the properties of these tunnelling processes are sensitive to the size of the system. In different words, the additional energy associated with these tunnelling processes is different for systems with different sizes and shapes.
     A direct manifestation of this sensitivity (when it is generalized to 4d case as we discuss below) is the emergence of the Topological Casimir Effect (TCE), when the vacuum energy and pressure depend on the size of the compact manifold on which the theory is defined. 
     
     Is this extra energy  physical? Our ultimate answer is ``yes". We refer to  Appendix \ref{2d} where we present some  arguments suggesting that  the extra energy  related to the tunnelling processes  in the ``empty" theory    can not be removed by any  redefinition of observables. It must be present in the system for consistency of the theory. In particular, the Ward Identities can not be maintained  without these tunnelling contributions, see Appendix \ref{2d} for details. Essentially, this extra contribution is precisely the source of violation of  a  commonly accepted (but   generally   wrong) receipt that the Casimir effect due to Maxwell photons could be obtained  by multiplying the corresponding scalar expressions by a factor of two.

  \section{Topological Casimir Effect in QED in four dimensions}\label{4d}
The topological structure of the gauge field in 2d can be easily generalized to higher dimensions. In four spacetime dimensions, we can devise boundary conditions that give rise to very similar instanton-like configurations, with precisely the action (\ref{action}) as found in 2d Maxwell theory on the torus. In this section we show that these topological degrees of freedom are completely decoupled from the propagating physical photons. Furthermore, the corresponding  quantum fluctuations  do not depend on the properties of the topological sectors (due to the linearity of the Maxwell equations), and can be treated in the conventional way. As a result we are able to focus on the new contributions and compare them to the conventional Casimir effect from literature. 
We shall see that the topological Casimir effect is strongly suppressed on a Euclidean 4-torus where one of the spatial dimensions is much smaller than the others. In this case the well known formulae (\ref{energy}) and (\ref{pressure}) are recovered. However, the main goal of this work is to study precisely those novel contributions which are sensitive to the system size.

\subsection{Decoupling of the topological and conventional parts}

To construct a theory defined on a Euclidean 4-torus\footnote{A Euclidean 4-torus in this case corresponds to a spatial 3-torus at finite temperature. This method, which corresponds to the Matsubara formalism, is a common way of calculating the Casimir effect at finite temperature.}, we consider a system with a box of sizes $L_1 \times L_2 \times L_3 \times \beta$ in the respective directions. A torus is realized when we assume periodic boundary conditions on the physical fields in all directions, in which case we find a ``degeneracy" of the vacuum state, just like in section \ref{topology}: by making  a loop in the $xy$-plane, the $A^{\mu}$ field can pick up a phase corresponding to a large gauge transformation. Working in Euclidean space and adopting the Lorentz gauge, it is simple to find a 4d generalization of the instanton potential from section \ref{pathintegral}  that satisfies these boundary conditions.
The 4d instanton potential is given by
\be
\label{toppot4d}
A^{\mu}_{top} = \left(0 ,~ -\frac{\pi k}{e L_{1} L_{2}} x_2 ,~ \frac{\pi k}{e L_{1} L_{2}} x_1 ,~ 0 \right),
\ee  
where $k$ is the winding number that labels the topological sector, and $L_{1}$, $L_{2}$ are the dimensions of the plates in the x and y-directions respectively, which are assumed to be much larger than the distance between the plates $L_3$. This classical configuration satisfies the periodic boundary conditions up to a large gauge transformation,  and provides a topological magnetic flux in the z-direction:
\be
\label{topB4d}
\vec{B}_{top} = \vec{\nabla} \times \vec{A}_{top} = \left(0 ,~ 0,~ \frac{2 \pi k}{e L_{1} L_{2}} \right) ,
\ee
in close analogy with the 2d case (\ref{inst}).
The Euclidean action of the system becomes
\be
\label{action4d}
\frac{1}{2} \int \dd^4 x \left\{  \vec{E}^2 +  \left(\vec{B} + \vec{B}_{top}\right)^2 \right\} ,
\ee
where the integration is over the Euclidean torus $L_1\times L_2\times L_3\times\beta$ and 
$\vec{E}$ and $\vec{B}$ are the dynamical quantum fluctuations of the gauge field.   These terms were not present in the 2d model, but must here be taken into account due to the presence of real propagating physical photons. We find that the action can be easily split into the sum of a topological and a quantum part, because of the vanishing cross term
\be
\int \dd^4 x~ \vec{B} \cdot \vec{B}_{top} = \frac{2 \pi k}{e L_{1} L_{2}} \int \dd^4 x~ B_{z} = 0 
\label{decouple}
\ee
Here the fact is used  that the magnetic portion of quantum fluctuations in the $z$-direction, represented by $B_{z} = \partial_{x} A_{y}  - \partial_{y} A_{x} $, is a periodic function because   $\vec{A} $ is periodic over the domain of integration. As a result, there is no coupling between  the conventional quantum fluctuations described by photons with physical polarizations and the classical instanton potential (\ref{toppot4d}), (\ref{topB4d}).  Furthermore, the  quantum fluctuations due to photons are not sensitive to the topological sector $k$  of the theory, and therefore they decouple from the classical $k$-instanton contribution. 
Finally, the  quantum fluctuations from photons must be computed in a box with size $L_1 L_2$ rather than in infinite space along $x, y$ directions. 
The corresponding corrections,   in principle, can be computed. Technically the computations would be quite tedious as they require the operation with Green's functions defined on a finite manifold rather than in the infinite space. The computations can be performed for the trivial $k=0$ topological sector as the corresponding corrections are independent of $k$. These contributions are expected to produce some corrections  
$\sim (1+\frac{a^2}{L^2})$ to formula (\ref{pressure}).  However, we shall not  elaborate on these  terms in the present work. It is a part of the conventional partition function  ${\cal{Z}}_{0}$ computed for trivial topological sector $k=0$.

 The main lesson from the previous discussion is that the conventional quantum fluctuations are not sensitive to the topological sectors $k$ as a result of linearity of the Maxwell equations. Therefore, they can be treated in completely separate ways, which greatly simplifies our analysis. For the partition function we can now write: ${\cal{Z}} = {\cal{Z}}_{0} \times {\cal{Z}}_{top}$. 	 The conventional part ${\cal{Z}}_{0}$ is well-studied for toroidal BCs at finite temperatures \cite{review}. It is $k$-independent,  so it will not be elaborated here. In the rest of this section we will study the behaviour of ${\cal{Z}}_{top}$.
 We shall see that in the limit when $\frac{L_1L_2}{L_3}\rightarrow\infty$ the partition function related to the topological effects yield ${\cal{Z}}_{top}=1$, and we recover the conventional Casimir effect (\ref{energy}), (\ref{pressure}) which is computed from  ${\cal{Z}}_{0}$.
 However, we shall see that a number of novel and unusual features will emerge in the system when $L_1, L_2$ are large but  remain finite, which    is precisely the main subject of our studies.

\subsection{Computing the topological pressure} \label{sec:toppressure}
The system of parallel plates is related to 2d Maxwell theory by dimensional reduction: taking a slice of the 4d system in the xy-plane will yield precisely the topological features of the 2d torus. Assuming that $L_3$ is much smaller than $L_1$ and $L_2$, the additional dimensions do not contribute toward ${\cal{Z}}_{top}$ as we noted above. Instead, the quantum corrections  slightly modify ${\cal{Z}}_{0}$ as they do not depend on the topological sectors $k$, and can be factored out  from ${\cal{Z}}_{top}$.
    With this set up,  the classical action  for configuration (\ref{topB4d}) takes the form 
\be
\label{action4d2}
\frac{1}{2}\int \dd^4 x \vec{B}_{top}^2= \frac{2\pi^2 k^2 \beta L_3}{e^2 L_1 L_2}
\ee
while  the topological partition function becomes:
\be
\label{Z4d}
{\cal{Z}}_{top} = \sqrt{\frac{2\pi \beta L_3}{e^2 L_1 L_2}} \sum_{k\in \mathbb{Z}} e^{-\frac{2\pi^2 k^2 \beta L_3}{e^2 L_1 L_2} },
\ee
where the 2d electric charge entering  eqs. (\ref{action}), (\ref{Z_4}) is expressed in terms of the 4d electric charge as follows\footnote{Note that the units are consistent. Indeed,  in 4d, $e^2\sim \alpha$ is the dimensionless fine-structure constant. However, in 1+1 dimensions the QED coupling constant has units of $(length)^{-2}$.}, 
\be
\label{dreduction}
e^2_{2d} = \frac{e^2}{\beta L_3}, ~~~  \frac{e^2}{4\pi}\equiv  \alpha.
\ee
In this section we consider $\theta=0$. More discussion on this matter is found in section \ref{B}.  

One should note that the dimensional reduction which is employed here is not the most generic one.
In fact, one can impose a non-trivial boundary condition on every slice in the 4d torus. However, the main goal of this work is not to classify the  most generic BC, but to discuss the physical properties for the simplest possible case (\ref{topB4d}), i.e. when a nontrivial BC is imposed on a single slice, while keeping  the  trivial periodic BC for other slices. 
  
With this objective in mind,  it is useful to introduce the dimensionless parameter
\be
\label{tau}
\tau \equiv {2 \beta L_3}/{e^2 L_1 L_2}
\ee
such that the partition function ${\cal{Z}}_{top}$ can be written in a very simple form:
\be
\label{Z_top}
{\cal{Z}}_{top} (\tau)= \sqrt{\pi \tau} \sum_{k\in \mathbb{Z}} e^{-\pi^2 \tau k^2} = \sum_{n\in \mathbb{Z}} e^{-\frac{n^2}{\tau}}, 
\ee
where  the Poisson summation formula (\ref{poisson}) is used again.
Our normalization  of the partition function ${\cal{Z}}_{top}$ 
is such that in the limit $L_1L_2=\infty$ the topological portion of the partition function ${\cal{Z}}_{top}
= 1$ so that we  
 recover the conventional Casimir effect (\ref{energy}), (\ref{pressure}) which is encoded in   ${\cal{Z}}_{0}$.
 The simplest way to check our normalization is to take the limit $\tau\rightarrow 0$ using  the right hand  side of eq. 
 (\ref{Z_top}) when a single term with $n=0$ contributes. It corresponds to very large instanton numbers $k\sim  {\tau}^{-1/2}\rightarrow\infty$  saturating  the original series (\ref{Z4d}).

From ${\cal{Z}}_{top}$, we can calculate any thermodynamic property of the system, like the topological pressure between the plates
\be
\label{toppressure}
P_{top} = \frac{1}{\beta L_1 L_2} \frac{\partial}{\partial L_3} \ln {\cal{Z}}_{top}.
\ee
In the asymptotic limit where $\tau\ll 1$ one can use the dual representation of ${\cal{Z}}_{top}$ encoded  by the Poisson re-summation formula (\ref{Z_top})  to find that
\be
\label{asymptotic}
P_{top} \approx \frac{e^2}{\beta^2 L_3^2} e^{-\frac{1}{\tau}} ~,~~~~ \tau \ll 1.
\ee
In this case, the topological pressure is exponentially suppressed, and when compared with the conventional Casimir pressure (\ref{pressure}) it is clear that the topological effect is too small to measure 
experimentally\footnote{The conventional Casimir pressure in eq. (\ref{pressure}) does not account for the thermal correction, and is computed for a system with slightly different boundary conditions (metallic instead of periodic). However, the boundary conditions only change the pressure by a constant of order unity and the thermal correction is negligible for low temperatures, see \cite{review}. Also note that even though $\tau \ll 1$, we are still in the low-temperature regime.}. 

A few comments are in order. 
First of all, one can explicitly see that  the original instanton formula (\ref{Z4d}) is consistent with our interpretation: that the additional energy and pressure due to the topological features of the system is the result of tunnelling events between different winding states, as discussed in section \ref{Interpretation}. Indeed, a non-analytical  structure of eq. (\ref{Z4d}) with respect to coupling constant $\exp(-1/{e^2})$ represents the typical behaviour for a  tunnelling process. It is quite fortunate that the Poisson re-summation formula (\ref{Z_top}) allows us to analyze both regimes, at large as well  as small $\tau$. 
 Secondly, even in this simple $\tau \ll 1$ case one can explicitly see that the sign of the effect is opposite to the conventional Casimir effect (\ref{pressure}). This correction  leads to  repulsive rather than  attractive forces. This ``wrong" sign is a typical manifestation 
of the topological fluctuations, in contrast with the conventional vacuum fluctuations of photons with physical polarizations. 
\exclude{
This ``wrong sign" of the effect is a result of
tunnelling events, not related to propagation and fluctuations of the real dynamical degrees of freedom. }See some additional comments on a ``wrong sign" in Appendix \ref{2d}. 

\begin{figure}[htb]
  \center{\includegraphics[width=\textwidth /2 ]{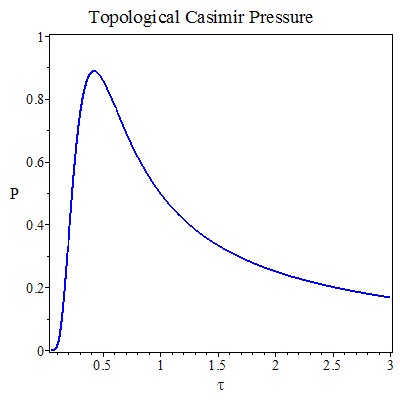}}
  \caption{\label{fig:taupressure}  The topological pressure on the 4d system of parallel plates as a function of $\tau \equiv {2 \beta L_3}/{e^2 L_1 L_2}$. Pressure is measured in units  $\frac{2}{L_1^2L_2^2e^2}$.}
\end{figure}

While $\tau \ll1$ can be examined analytically, it is more interesting to study  a system with $\tau \simeq 1$ where this effect could be sufficiently large. We can satisfy the conditions $L_1, L_2\gg \beta, L_3$, in order to use the dimensional reduction, and at the same time still achieve $\tau \simeq 1$ because 
    the small parameter $e^2$ enters the denominator in eq. (\ref{tau}) in the definition for $\tau$.
 In this case we have to resort to numerical approximations, since there is no closed form for the partition sum and pressure. In Figure \ref{fig:taupressure}, a numerical plot of $P_{top}$ is shown for this regime. There is a large peak around $\tau \simeq 0.4$ where the pressure, measured in units $\frac{2}{L_1^2L_2^2e^2}$, has an order of 1. 
The relative magnitude between the maximum topological pressure and the conventional Casimir pressure (\ref{pressure}) using parallel ideal conductors at low temperature is thus approximately 
\be
\label{ratio}
R_{max}\approx \frac{|P_{top}|}{|P|}\approx \frac{480L_3^4}{L_1^2L_2^2e^2\pi^2}\approx\frac{120}{\pi^3\alpha}\cdot\frac{L_3^4}{L_1^2L_2^2}.
\ee
This ratio (even at its maximum at $\tau \simeq 0.4$) is very small in a typical Casimir experiment setup with $L_1, L_2 \gg L_3$,   in spite of the large numerical factor in front of formula (\ref{ratio}). As we mentioned previously, the power-like corrections $\sim L_3^2/L_1^2,  L_3^2/L_2^2$ are also expected to occur  in    ${\cal{Z}}_{0}$ resulting  from the conventional vacuum fluctuations of physical photons. We expect these conventional corrections to be even smaller as they can not contain a parametrically enhanced factor $1/e^2$ that is a unique feature of the topological vacuum fluctuations.

To conclude this  section, we find that there is a small, but very real, contribution to the Casimir effect that is purely due to topological features of the system. When QED is defined on a compact manifold such as a 4-torus, one needs to take into account the tunnelling processes which occur  between the topologically inequivalent (but physically identical) winding states. These topological transitions are described in terms of integer magnetic fluxes (\ref{topB4d}). It is not surprising that the effect is exponentially small in normal circumstances (\ref{asymptotic}). The effect remains very small (\ref{ratio}) even at $\tau\sim 1$. Still, there is a hope to make it  measurable by studying the Topological Casimir Effect (TCE) in the presence of some external   magnetic field. We shall observe  a high sensitivity of TCE to applied weak  external magnetic field. This should be contrasted   with conventional Casimir effect (\ref{energy}), (\ref{pressure}) which can not be sensitive to external fields as 
vacuum photon fluctuations  do  not couple to external fields (since the Maxwell equations are linear).   This topic is precisely the subject of the next section.

    \section{$\theta $ vacua and external magnetic fields}\label{B}

Now it is interesting to place our system into a region with a weak external magnetic field $B^{ext}_z$ along the $z$-direction.
The idea behind this construction is that the external  magnetic field $B^{ext}_z$ will interfere with the integer topological flux (\ref{topB4d}) describing the tunnelling events. It is expected that such interference   may  skew the summation over the topological sectors, similar  to the action of the so-called $\theta$ parameter (\ref{Z_3}), (\ref{Z_4}). As we shall demonstrate below, this is indeed  what happens  in our simple case considered in section \ref{4d}. In different words, 
 we claim that by adding a constant magnetic field to the previous   setup, an effective  non-zero $\theta_{eff}$ parameter emerges in the system. The crucial point here is that we introduce this  parameter which can be externally varied. By studying the corresponding responses to $\theta_{eff}$ variation, it gives us some hope that while the TCE is numerically very small (\ref{ratio}),  it is nevertheless very sensitive to a weak magnetic external field 
 (in contrast with conventional Casimir effect (\ref{pressure})), and hopefully it can be eventually measured due to this sensitivity. 

To construct a system as such, in addition to the topological flux through the xy-plane, we apply a real physical constant magnetic field $B^{ext}_z = \partial_x A_y^{ext} - \partial_y  A_x^{ext}$ parallel to the $z$-direction (perpendicular to the $xy$-plane). The total $B_z$ field in the Euclidean metric is thus modified as follows:
\begin{dmath}
 B_z=   B^{q}_z+B^{top}_z+B^{ext}_z
\end{dmath}
where the total field decomposition consists of the same instanton potentials $A_{\mu}^{top}$ as in eq. (\ref{toppot4d}), the external magnetic field potential $A^{ext}_{\mu}$ given above and the quantum fluctuations $A^q_{\mu}$ around them. 
\exclude{The modified Euclidean action of this system is denoted by $S_{\theta_{eff}}[A^{(k)}]$, where $k$ corresponds to the $k$-th topological sector, while $\theta_{eff}$ parameter is defined below by eq. (\ref{theta_eff}).}

The only difference from the previous construction is the additional external constant magnetic field. Note that  the quantum fluctuations still decouple from the classical and external fields, similar to    eq. (\ref{decouple}),    due to the periodicity of quantum fluctuations over the domain of integration, 
\be 
 \Delta S 
= (B_z^{ext}+\frac{2\pi k}{L_1L_2e})\cdot\int d^4 x  B^{q}_z=0.
\label{eqn:decouplingExtB}
\ee
The remaining part of the action  is quadratic and thus path integration can be performed. The same calculation from the previous section follows and the partition function separates into a classical portion, which describes TCE, and a quantum portion that corresponds to the effect of photons in 4D, i.e. the well known Casimir effect.
It is important  that the  conventional quadratic term representing the    photon fluctuations  does not depend on the topological sector $k$, nor on the external magnetic field. It is described exclusively by ${\cal{Z}}_0$, as before.

The new element here is  that the external field   couples  to  the instanton potential.
Therefore, the topological partition function now takes the form, 
\be 
\label{Z_eff}
  {\cal{Z}}_{top}(\tau, \theta_{eff})
 =\sqrt{\pi\tau} \sum_{k \in \mathbb{Z}} \exp\left[-\pi^2\tau \left(k+\frac{\theta_{eff}}{2\pi}\right)^2\right]  ~~
\ee
where we introduced the effective theta parameter 
\exclude{\footnote{Note that the effective theta parameter $\theta_{eff}$ is not exactly the same as the $\theta$ that was typically defined in a theta state, although they are very similar}}
\be
\label{theta_eff}
\theta_{eff} = B^{ext}_z L_1L_2e
\ee
proportional to the external magnetic flux through the xy-plane in this particular system. It is clear from the partition sum 
 (\ref{Z_eff})
that a non-zero effective $\theta_{eff}$ skews the summation over topological sectors similar to the 2d example given by (\ref{Z_3}), (\ref{Z_4}).  It is also clear that 
$\theta_{eff}=2\pi m$ corresponds to integer flux $m$ through the xy-plane,  which obviously can not modify the system, such that   ${\cal{Z}}_{top}(\tau, \theta_{eff})$  is $2\pi$ periodic in $\theta_{eff}$.

In what follows, we also need a ``dual" representation for ${\cal{Z}}_{top}(\tau, \theta_{eff})$ which is obtained by applying the Poisson re-summation formula (\ref{poisson})
\be
\label{Z_dual}
  {\cal{Z}}_{top}(\tau, \theta_{eff})
 = \sum_{n \in \mathbb{Z}} \exp\left[ -\frac{n^2}{\tau}+i ~\theta_{eff}  n \right].
\ee
In representation (\ref{Z_dual}), it is obvious that $\theta_{eff}$ being  expressed in terms of the external magnetic field (\ref{theta_eff})  can be thought of as a fundamental $\theta$ parameter. However, the corresponding ``instanton charge" $n$  which normally enters with $\theta$ is not the same magnetic flux $k$
from  our original construction (\ref{topB4d}) with classical action (\ref{action4d2}). Rather, it is some ``dual" configuration with classical action $\sim \tau^{-1}$. 

We note that $ {\cal{Z}}_{top}(\tau, \theta_{eff})$ is properly normalized    in the limit $L_1, L_2\rightarrow\infty$ which corresponds to $\tau\rightarrow 0$. In this case, only a single term with $n=0$ in (\ref{Z_dual}) survives, leading to the desired normalization $ {\cal{Z}}_{top}(\tau\rightarrow 0, \theta_{eff})=1$. Therefore, all conventional formulae for the Casimir effect determined by $ {\cal{Z}}_{0}$ are recovered in this limit as $ {\cal{Z}}$
is factorized $ {\cal{Z}}={\cal{Z}}_{0}\times {\cal{Z}}_{top}$ for our system. 

Now we are in position to calculate the topological Casimir pressure contribution from free energy at finite temperature, by inserting the partition function into eq. (\ref{toppressure}).
Like in section \ref{sec:toppressure}, the topological pressure has no closed form. For the limit where $\tau\ll 1$, we may obtain an asymptotic expansion. Using the dual representation (\ref{Z_dual}) and keeping the leading order terms, we arrive at:
\be 
P_{top} \approx \frac{e^2}{\beta^2L_3^2} \cos(\theta_{eff}) \exp(-1/\tau) , ~~ \tau\ll1
\ee
which reduces to our previous formula (\ref{asymptotic}) in the absence of the external magnetic field.  
As expected, the oscillatory effect with respect to $\theta_{eff}$   is present, but becomes exponentially suppressed because the tunnelling amplitudes naturally diminish in this limit.

\begin{figure}[htb]
  \center{\includegraphics[width=\textwidth /2 ]{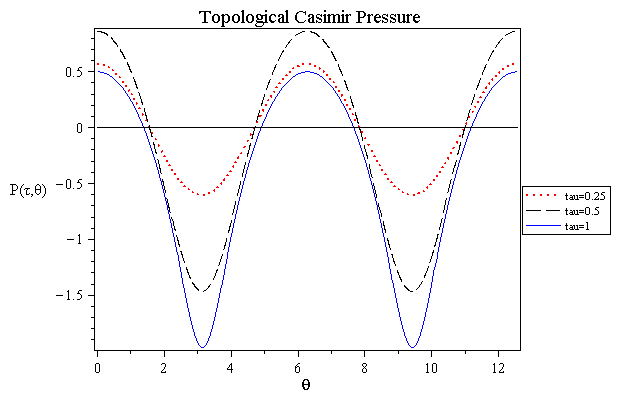}}
  \caption{\label{fig:pressure}   Topological Casimir   Pressure is plotted in units $\frac{2}{L_1^2L_2^2e^2}$ for different $\tau$. A clear $2\pi$ periodicity is seen and local extrema are between odd and even integer multiples of $\pi$. }
\end{figure}

In a more general  case when  $\tau\simeq 1$,  we have to use some numerical methods
as  asymptotic analysis is no longer sufficient. A numerical plot is shown in Figure \ref{fig:pressure}, such that the variation with $\theta_{eff}$ is manifest. The  pressure is clearly oscillatory with respect to $\theta_{eff}$ and its local extrema are attained at $n \pi$ where $n\in\mathbb{Z}$. Thus, by altering the magnetic flux, the topological Casimir pressure will also be modified accordingly. Additionally,  at $\theta_{eff}=0$ the pressure has a ``wrong" sign, i.e. it is opposite to conventional Casimir effect (\ref{pressure}),  as we already discussed after eq.(\ref{asymptotic}).    This sign changes as a function of the external magnetic field, as can be seen in the plot. Such a variation can be interpreted as the result of interference between the external magnetic field and the topological ``instanton" fluctuations (\ref{topB4d}).
This very specific and distinct variation, hopefully,  might be a useful feature  to measure the effect in the future in spite  of its strong numerical suppression (\ref{ratio}) in comparison with the conventional Casimir effect(\ref{pressure}).  

Our next step  is to analyze 
the magnetic response of the system under influence of the external magnetic field.  The idea behind these studies is the observation that the external magnetic field acts as an effective $\theta_{eff}$ parameter as eqs. (\ref{theta_eff}), (\ref{Z_dual}) suggest. Therefore, the corresponding study of 
the magnetic response in our system is in fact  very similar to an analysis of the expectation value of the topological density and topological susceptibility in other gauge models, where a $\theta$ parameter enters the system in a similar way. See e.g. Appendix \ref{2d} for some discussion on this in 2d Maxwell theory. 

First, we compute  
 the induced magnetic field defined as follows
\be 
\label{B_ind}
&\,& \langle B_{ind} \rangle = -\frac 1 {\beta V}\frac{\partial \ln \mathcal{Z}_{top}}{\partial B^{ext}_z}=
-\frac{e}{\beta L_3}\frac{\partial \ln\mathcal{Z}_{top}}{\partial\theta_{eff}}\\
&=& \frac{\sqrt{\tau\pi}}{\mathcal{Z}_{top}}\sum_{k\in\mathbb{Z}}\left(B^{ext}+\frac{2\pi k}{L_1L_2e}\right)\exp{\left[-\tau\pi^2(k+\frac{\theta_{eff}}{2\pi})^2\right]}.\nonumber
\ee
As one can see from (\ref{B_ind}),  our definition of the induced field  accounts for the total field which includes both terms: the external part as well as the topological portion of the field. In the absence of the external field ($B^{ext}=0$), the series is antisymmetric under $k\rightarrow -k$ and $\langle B_{ind} \rangle$ vanishes. It is similar to the vanishing expectation value of the topological density  in gauge theories  when $\theta=0$. One could anticipate this result from symmetry arguments as the theory must respect $\cal{P}$ and $\cal{CP}$ invariance at $\theta=0$.

The expectation value of the induced magnetic field exhibits similar $2\pi$ periodicity from the partition function and it reduces to triviality whenever the amount of skewing results in an antisymmetric summation, i.e. $\langle B_{ind}\rangle = 0$ for $\theta_{eff}\in\{n\pi:n\in\mathbb{Z}\}$. This property is analogous to the well known property in QCD when $\cal{P}$ and $\cal{CP}$ invariance holds only for $\theta={n\pi}$.

\begin{figure}[htb]
  \center{\includegraphics[width=\textwidth /2 ]{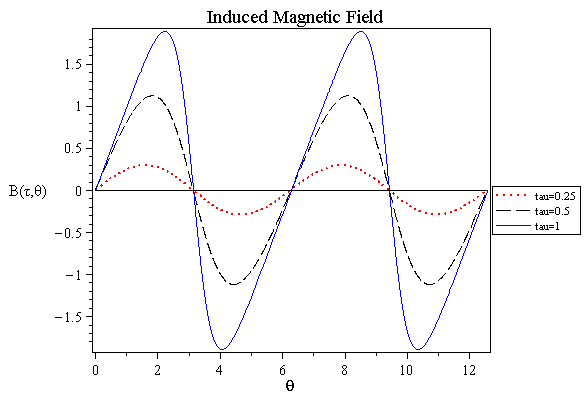}}
  \caption{\label{fig:bfield} A numerical plot  of the induced magnetic field in units  $\frac{1}{L_1L_2e}$ as a function of $\theta_{eff}$.   The oscillatory behaviour becomes more pronounced  for large $\tau$.}
\end{figure}

As before, the topological  effects are exponentially suppressed at $\tau\ll 1$, as $ \mathcal{Z}_{top}\rightarrow 1$ with exponential accuracy at $\tau\ll 1$. The effect  is much   more pronounced in the range where $\tau\simeq 1$, see Figure \ref{fig:bfield}, where we plot  the induced magnetic field in units  $(L_1L_2e)^{-1}$ as a function of $\theta_{eff}$.   One should also remark here that the induced magnetic field defined as (\ref{B_ind}) can be thought of as the magnetization of the system per unit volume, i.e. $\la M\ra=-\la B_{ind}\ra$, as the definition for 
$\la M\ra$ is identical to (\ref{B_ind}) up to a minus sign.

Now we turn our attention to the magnetic susceptibility, which is similar to the topological susceptibility reviewed in Appendix \ref{2d} for 2d QED. This object is $\cal{P}$ and $\cal{CP}$ even and does not   vanish at zero external field. The magnetic susceptibility measures the response of free energy to the introduction of a source term,
which is represented in our case by $B^{ext}\sim \theta_{eff}$.
To be more precise, we define $ \chi_{mag}$ in a way which is similar  to the topological susceptibility in Appendix \ref{2d} for 2d QED, 
\be 
 \chi_{mag}=  \int d^4x \la B_z(x), B_z(0)\ra=  -\frac{1}{\beta V} \frac{\partial^2 \ln \mathcal{Z}_{top}}{\partial B_{ext}^2}, 
\label{definition}
\ee
 where the integration is over the Euclidean torus $L_1\times L_2\times L_3\times\beta$. With this definition, $\chi_{mag}$ is a  dimensionless parameter, in contrast with 2d QED where $\chi_{E\&M}$ has dimension $(mass)^2$, and in 4d QCD where $\chi_{QCD}$ has dimension $(mass)^4$. This is due to the fact that the topological density operator 
 has dimension $(mass)^2$ in 2d QED and $(mass)^4$ in 4d QCD while our topological instanton (\ref{topB4d}) expressed in terms of the magnetic field has dimension $(mass)^2$. Nevertheless, we opted to keep definition (\ref{definition}) without inserting any additional dimensional parameters such as $L_1L_2$ into formula  (\ref{definition}), to maintain the conventional definition in statistical mechanics where $\chi_{mag}$ is a dimensionless parameter  (in units where $\hbar=c=1$). 
 
\begin{figure}[htb]
  \center{\includegraphics[width=\textwidth /2 ]{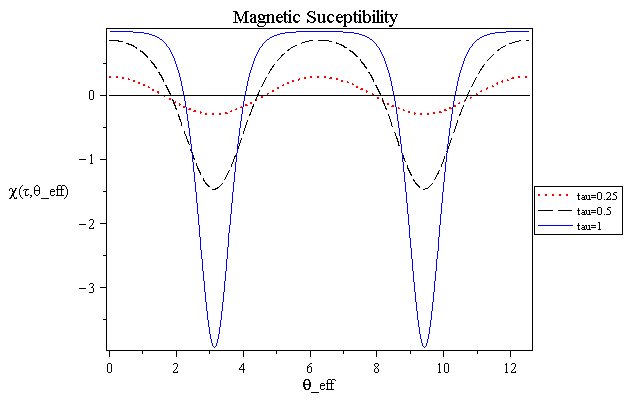}}
  \caption{\label{fig:suscept} Numerical plot of magnetic susceptibility $\chi_{mag}$ as a function of $\theta_{eff}$ for different values of $\tau$. It oscillates with $\theta_{eff}$ as it should,  and it does not vanish even  at zero external magnetic field at $\theta_{eff}=0$ due to topological fluctuations, similar to well studied cases of 2d QED and 4d QCD.}
\end{figure}

 We can represent (\ref{definition}) in terms of  dimensionless variables as follows,
 \be 
 \chi_{mag} = -\frac{2}{\tau} \frac{\partial^2 \ln Z_{top} (\tau, \theta_{eff})}{\partial \theta_{eff}^2}.
\label{eqn:suscept}
\ee
In the limit when $ \tau \ll 1$, one can use analytical expression (\ref{Z_dual}) to conclude   that $\chi_{mag}\sim \exp(-\frac{1}{\tau})$ is strongly suppressed. 
It is consistent with our expectations that there should not be any magnetic correlations in the conventional Casimir experimental setups. However, 
for $\tau$ near the order of 1, the behaviour is quite nontrivial as shown in Figure \ref{fig:suscept}.
Note that in this case, the susceptibility is highly dependent on the external field and changes signs, which is extremely unusual for conventional systems. More specifically, $\chi_{mag}$ behaves like a weak diamagnetic for $\theta_{eff}\approx\pi n$, $n$ even; and behaves as strongly paramagnetic for $n$ odd.

We note that this topological effect  is quite distinct from the behaviour of the conventional Casimir effect. In the conventional quantization of electromagnetic fields in Minkowski space, there is no \emph{direct} connection  between the Casimir pressure and an external magnetic field\footnote{This holds for the present assumption where radiative corrections are not considered and the boundary conditions are not affected by the magnetic field. Note that in the case of the TCE it is the \emph{bulk} vacuum that is sensitive to an external magnetic field.}. Although such coupling could be attained through fermionic interactions in higher order diagrams, it will be highly suppressed $\sim \alpha^2B_{ext}^2/m_e^4$ as the non-linear Euler-Heisenberg Effective Lagrangian would suggest. In fact the explicit computations of this effect have been done \cite{Robaschik:1987} and they fully agree with our order of magnitude  estimates\footnote{ We are thankful to our anonymous referee for pointing out that the corresponding computations have in fact been explicitly performed in \cite{Robaschik:1987} for the case of conventional Casimir effect}. Thus, in a pure Maxwell theory as we consider here, the photon-photon interactions should be trivial. However, in TCE, we see that the external magnetic field does have a non-trivial effect on physical quantities through its interactions with topological ``instanton" fluctuations (\ref{topB4d}). To put this in more concrete terms, the numerical value for loop level corrections in the conventional Casimir effect is of order $~10^{-20}$  even with a 1T external magnetic field. In contrast, the proposed correction to TCE in an external field is of order 1, as shown in Figure \ref{fig:pressure}. 

Therefore, the periodicity in all of the physical quantities with respect to the external magnetic field is a unique feature of TCE. It is not found in any of the typical Casimir results and can serve as a clear indicator to distinguish a topological effect from conventional Casimir effects.

    \section{Conclusion and future directions}\label{conclusion}

We have demonstrated the existence and properties of a new type of vacuum fluctuations in gauge fields, resulting from the summation over topological sectors.
While most literature on the Casimir effect neglects these topological sectors, which are indeed absent in the topologically trivial Minkowski space, they need to be taken into account when  the  theory is formulated on a non-simply connected, compact manifold.
 Physics related to pure gauge configurations describing the  topological sectors  does not go away when one removes all unphysical degrees of freedom as a result of gauge fixing; instead, this physics  reappears in a much more complicated form where the so-called Gribov ambiguities\footnote{A short historical remark is warranted here.  The Gribov ambiguities \cite{Gribov:1977wm} were originally discussed for non-abelian gauge theories  in Minkowski space when one tries to completely remove all unphysical degrees of freedom in the Coulomb gauge. The corresponding  Gribov ambiguities lead to strong infrared singularities in the consequent analysis. Another option is to deal explicitly with some unphysical degrees of freedom that effectively describe these topological sectors.} emerge \cite{Gribov:1977wm}. See recent paper \cite{Kelnhofer:2012ig} and also some previous relevant discussions \cite{Killingback:1984en,Parthasarathy:1988sa,Kelnhofer:2007zk}. We opted to keep some gauge freedom in our analysis to 
study these topological sectors explicitly. 
  
  Now we can formulate the main results of the present paper. First of all, the physics behind the Topological Casimir Effect (TCE) is quite simple: there are tunnelling events in a theory formulated on a  small compact manifold when the  temperature is small but not identically zero. 
  These transitions can be completely ignored for relatively  large systems, but in general, these topological transitions interpolating between different winding states (which correspond to one and the same unique physical state)  do occur. 
  The amplitudes for these transitions depend on size and shape of the system. Therefore, it is not really a surprise that the vacuum energy associated with these tunnelling events depends on the size of the system, which ultimately implies an extra contribution to the Casimir pressure. 
  
  In general, the effect is numerically much smaller than the conventional Casimir effect with the ratio given by eq.(\ref{ratio}). However, we argued that the effect is highly sensitive to small external magnetic fields which can serve as a clear indicator to distinguish TCE  from conventional Casimir effects.
 
 Our last comment is as follows. The TCE as we already mentioned in the introduction is a very generic phenomenon in gauge theories. It shows the algebraic sensitivity to the size of the system even when the theory has a mass gap. See \cite{Zhitnitsky:2011tr,Zhitnitsky:2011aa} and many references therein where TCE has been tested in various models, including QCD lattice computations. Our comment here is that the observed Dark Energy (DE) in the universe might be a direct manifestation of the TCE as argued in  \cite{Zhitnitsky:2011tr,Zhitnitsky:2011aa} and references therein. The idea is based on two key elements. Firstly, the additional energy in Maxwell theory (defined on a compact manifold and discussed in this paper) is based on  nontrivial topological properties formally expressed by the first homotopy group 
 $\pi_1 [U(1)] \sim \mathbb{Z}$. In four dimensions a similar structure emerges  for  non-abelian QCD where  the third homotopy group is non-trivial $\pi_3 [SU(3)] \sim \mathbb{Z}$. In this case one can argue that the  system is   algebraically  sensitive  to very  large distances in spite of the fact that the theory has a mass gap.
The second key element is based on   the paradigm that the relevant definition of  the energy  which   enters  the Einstein equations 
 is the difference $\Delta E\equiv (E -E_{\mathrm{Mink}})$, similar to the   Casimir effect (\ref{energy}), rather than the energy $E$  itself. 
 In this case the difference between the two metrics (expanding universe with Hubble expansion rate $H$   and Minkowski space-time) as a result of TCE would lead to an estimate  \cite{Zhitnitsky:2011tr,Zhitnitsky:2011aa} 
   \be
   \label{Delta}
   \Delta E\sim L^{-1}\Lambda_{\mathrm{QCD}}^3\sim (10^{-3} {\text eV})^4,
   \ee
 where  $L$ is the visible size of the universe estimated as $L^{-1}\sim H\sim 10^{-33} {\rm eV}$. Estimation (\ref{Delta})  
  is amazingly close to the observed DE value today. In fact, 
  a comprehensive phenomenological analysis of this model (the so-called ``ghost dark energy" model)  has been recently studied in a number of papers where comparisons have been made with the current observational data. (See  references on observational papers in  \cite{Zhitnitsky:2011tr,Zhitnitsky:2011aa}.)  The conclusion  was that the  model (\ref{Delta}) is consistent with all presently available data, and we refer the reader to the original  papers on analysis of the observational data.

  Our comment relevant for the present study is that some very fascinating topological properties of the quantum vacuum 
  may be, in principle, studied in a laboratory if the TCE in Maxwell theory, which is the main subject of the present work, can be experimentally measured. We conclude on this optimistic note\footnote{In fact, the idea to test some intriguing vacuum properties relevant for cosmology in a laboratory is not a very new idea. It has been advocated by G. Volovik  for years,   see recent review \cite{Volovik:2011dy} and references therein.}.

\section*{Acknowledgements}
One of authors (ARZ) is thankful to the anonymous referee who reviewed  ref.\cite{Zhitnitsky:2011tr} for a question which actually motivated the present work. Another author (C.Cao) also wishes to thank NSERC for the USRA fellowship that helped fund this project.
This research was supported in part by the Natural Sciences and Engineering Research Council of Canada.

\appendix

\section{Why topological vacuum fluctuations must be real and physically observable.}\label{2d}
The main subject of the present work is zero point (vacuum)  fluctuations. There has always hung a shadow over this question as there have always been suspicions that those vacuum fluctuations are not really  zero point fluctuations, but rather can be attributed to some other physics. See in particular the relatively recent paper  \cite{Jaffe:2005vp} where it has been argued that the conventional Casimir effect can be computed without even mentioning such a notion as the ``vacuum". 

The main goal of this appendix is to argue, using a simple exactly solvable 2d model, that the topological vacuum fluctuations are very real and very physical. In other words, we want to present a few arguments suggesting that finite contributions  resulting from   topological features of the system cannot be removed by any means such as subtraction or redefinition of observables. 
 
  We start our study with the topological susceptibility $\chi$   defined as follows,
  \be
\label{chi1}
\chi \equiv \frac{e^2}{4\pi^2} \lim_{k\rightarrow 0} \int \dd^2x ~e^{ikx}\left< T E(x) E(0) \right> ,
\ee
where $Q=\frac{e}{2\pi}E$ is the topological charge density 
and 
\be
\label{k}
\int  \dd^2x ~Q(x)= \frac{e}{2\pi} \int \dd^2x ~E(x) =k
\ee
is the integer valued topological charge   in the 2d $U(1)$ gauge theory, $E(x)=\partial_1A_2-\partial_2A_1$ is the field strength. 
The   $\chi$  measures response  of the free energy to the introduction of a source  term 
\be
\label{theta}
L_{\theta}=i \theta\frac{e}{2\pi}\int \dd^2x~ E(x).
\ee
The corresponding computations can be easily carried out  as the partition function ${\cal{Z}}(\theta)$ is known exactly, see section \ref{topology}. When differentiating the partition function twice with respect to $\theta$, we  get a finite contribution in the infinite volume limit, $V\equiv \beta L\rightarrow \infty$, i.e  
\be
\label{exact1}
\chi_{E\&M}= -\frac{1}{\beta L}\frac{\partial^2 \ln{\cal{Z}}(\theta)}{\partial\theta^2}|_{\theta=0}= \frac{e^2}{4\pi^2}.  
\ee
  Is    contribution (\ref{exact1}) physical? This question immediately arises because we are dealing with  ``empty" Maxwell theory in two dimensions, where there are no physical propagating degrees of freedom   in the system. Can we redefine the theory to 
 remove all such terms from consideration once and for all? For example, one can use a prescription \cite{Barvinsky:1995dp} which  ignores the topological sectors  and leads to a trivial partition function $ {\cal{Z}}=1$, see eqs (4.1), (4.3) in \cite{Barvinsky:1995dp}. 
 Such a prescription would obviously  be consistent with the conventional procedure which   relates the Casimir effect for 4d Maxwell theory and 4d massless scalar field theory up to  factor 2. However, such a prescription would not produce the contact term (\ref{exact1}) which must be present in the system for its consistency, as we shall argue below. 
 
 The question addressed  above on ``physical reality" of (\ref{exact1})  is not a purely academic question. If eq. (\ref{exact1})  is treated as a physical contribution, then the partition function ${\cal{Z}}(\beta, L,\theta) $ which leads to  (\ref{exact1}) is also physical.  Therefore, the same    partition function ${\cal{Z}}(\beta, L,\theta) $ will also lead to  an   extra Casimir force  $P\sim  \partial \ln{\cal{Z}}/\partial L$ which  ``in principle" is an observable quantity. In different words, if (\ref{exact1}) is physical, then there is an extra term in the  Casimir energy which is not related to any  asymptotic  propagating degrees of freedom and is ``in principle" observable. We present a few arguments below to advocate that (\ref{exact1}) is indeed physical.

 Our argument  goes as follows. We add a massless fermion  field $\psi$  to the system  to arrive at the well known 2d Schwinger model. 
 The  expression for the topological susceptibility in the 2d Schwinger QED model is known exactly \cite{SW,Zhitnitsky:2011tr}
 \be
\label{exact}
\chi_{QED}= \frac{e^2}{4\pi^2}  \int   \dd^2x \left[ \delta^2(x) - \frac{e^2}{2\pi^2} K_0(\mu |x|) \right] ,
\ee
where $\mu^2=e^2/\pi$ is the mass of the single physical state in this model, and $K_0(\mu |x|) $ is the modified Bessel function of order $0$, which 
is the Green's function of this   massive particle. The crucial observation here is as follows: any physical state contributes to $\chi_{QED}$ with negative sign 
\be
\label{dispersion}
\chi_{dispersive} \sim  \lim_{k\rightarrow 0} \sum_n  \frac{\la 0| \frac{e}{2\pi}   E  |n\ra \la n | \frac{e}{2\pi}   E |0\ra }{-k^2-m_n^2} <0.
\ee
In particular, the term proportional to $ -K_0(\mu |x|) $ with the negative sign  in eq. (\ref{exact}) is the result of the only physical field  of   mass $\mu$. 
However,   there is also a contact term $ \frac{e^2}{4\pi^2} $ in eqs. (\ref{exact}), (\ref{exact1}) which contributes to the topological susceptibility $\chi$  with the {\it opposite sign}, and which can not be identified according to (\ref{dispersion}) with any contribution from any   asymptotic state. 

The first term $ \frac{e^2}{4\pi^2} $ in this formula (\ref{exact})  can be easily recognized  as 
the expression for $\chi_{E\&M}$ for 2d Maxwell theory   (\ref{exact1}) which is originated    from  the  topological sectors, and not related to propagating degrees of freedom.
 
 This term has a fundamentally different, non-dispersive  nature. In fact it is ultimately related to different topological sectors  
 discussed in section \ref{topology}.
 This contact term must be present in the expression (\ref{exact}) to satisfy the Ward Identity (WI)  which states that $\chi_{QED}(m=0)=0$, see \cite{Zhitnitsky:2011tr} for the details.
   Without this   contribution, it would be impossible to satisfy the Ward Identity   because the physical propagating degrees of freedom can only contribute with sign $(-)$ to the correlation function as eq. (\ref{dispersion}) suggests, while WI requires $\chi= 0$  in the  chiral limit $m=0$.
 One can explicitly check that WI is indeed automatically satisfied only as a result of exact cancellation between conventional dispersive term with sign $(-)$ and non-dispersive term (\ref{exact1}) with sign $(+)$,   
 \be
\label{chi2}
\chi  &=& \frac{e^2}{4\pi^2}  \int   \dd^2x \left[ \delta^2(x) - \frac{e^2}{2\pi^2} K_0(\mu |x|) \right] \\  \nonumber
&=& \frac{e^2}{4\pi^2} \left[ 1- \frac{e^2}{\pi}\frac{1}{\mu^2}\right]= \frac{e^2}{4\pi^2} \left[ 1-1\right]=0.
\ee
 The lesson we learn from this simple exercise is that the contact term (\ref{exact1}) which is saturated by the  topological sectors is physical, and it must be present in the system for its consistency.
 
 The same contact  term (\ref{exact1}, \ref{exact}) can be also computed  using the auxiliary ghost fields, the so-called Kogut-Susskind (KS) ghost,  as has been originally done in ref. \cite{KS}, see \cite{Zhitnitsky:2011tr} for relevant  discussions in the present context. This auxiliary ghost field effectively takes into account the presence  of  topological sectors which lead to  (\ref{exact1}). 
 The crucial element accounting for  different topological sectors
of the underlying  theory,  does not go away    in KS-description. Rather,  this information is now coded  in terms of the
unphysical ghost scalar field which provides  the required ``wrong" sign for  contact  term (\ref{exact1}, \ref{exact}).
 The   contact term     in this framework is precisely represented 
by the ghost contribution  replacing the standard procedure 
of summation over different topological sectors.   
At the same time, this unphysical ghost scalar field  does not violate unitarity or any other important 
properties of the theory as  consequence  of Gupta-Bleuler-like condition on the physical Hilbert space, see \cite{Zhitnitsky:2011tr} for the details in the given context\footnote{It is important  to emphasize that the KS ghost should  not be confused with 
the conventional Fadeev-Popov ghost which is normally introduced into the theory
to cancel out unphysical polarizations of the gauge fields. Instead, the KS ghost is introduced to account for the  existence of  topological sectors
in the theory. A similar construction is also known for four dimensional non-abelian gauge theory where the corresponding color singlet field is called the Veneziano ghost, see \cite{Zhitnitsky:2011tr} for references and details. In the four dimensional case the Veneziano ghost can not be confused with a Fadeev-Popov ghost as the Veneziano ghost being a singlet does not carry  a color index, in contrast  with Fadeev-Popov ghosts.
 The sole purpose of the Veneziano ghost is to saturate 
 the contact term with the ``wrong sign" in the topological susceptibility, similar to eq. (\ref{exact1}), (\ref{exact}) in the 2d Schwinger model.   }.

 Our second argument that the topological sectors must be taken into consideration is based on analysis  of the entropy 
  in the same 2d ``empty" Maxwell theory. Before we formulate our argument, we want to make a short historical detour on the entropy studies  in this ``empty" model.  
  
  It has been claimed \cite{Kabat:1995eq} that for spins zero and one-half fields, the one loop correction to the black hole entropy is equal to the entropy of entanglement, while for a spin one Maxwell field, the entropy has an extra term describing the contact interaction with the horizon.  While the entropy is a positively defined entity, the Kabat contact term   is negative \cite{Kabat:1995eq}. Furthermore, this term being a total divergence can be represented as a surface term determined by the behaviour of the theory at arbitrarily large distances, i.e. it obviously has an infrared (IR) origin. More recently, it has been conjectured  \cite{Zhitnitsky:2011tr} that the  Kabat contact term is originated from the same topological gauge sectors which saturate the topological susceptibility (\ref{exact1}). 
  Indeed,  both terms have ``wrong" signs in comparison with what physical propagating degrees of freedom would produce, and both terms can be represented by surface integrals, see  \cite{Zhitnitsky:2011tr}  for the details. Next step in this development was the computation   of the entropy for   the 2d Maxwell system, defined on a finite dimensional compact manifold with
  size $V=\beta L$, such that the IR physics can be properly treated \cite{Donnelly:2012st}, see also \cite{Kabat:2012ns,Solodukhin:2012jh} with related discussions. In this case the expression for the entropy can be easily computed from the partition function (\ref{Z_2}), (\ref{Z_4}), (\ref{poisson}) discussed in section \ref{topology}, and is given by  \cite{Donnelly:2012st}
  \be
  \label{entropy}
  S=\left(\ln {\cal{Z}}+\frac{1}{2}\right)-\frac{1}{2}\left(\frac{4\pi^2}{e^2}\right)\cdot\chi_{E\&M},
  \ee
where $\chi_{E\&M}$ is the topological susceptibility given by eq. (\ref{exact1}). One can explicitly see that  the  negative contribution is indeed present in  the expression (\ref{entropy}) for the entropy.   This term with the ``wrong" sign in   eq.(\ref{entropy}) is exactly proportional to the  topological susceptibility (\ref{chi1})  in agreement with conjecture \cite{Zhitnitsky:2011tr}.  
Furthermore, this term    can be represented as   a surface integral because  $Q=\frac{e}{2\pi}E$ entering (\ref{chi1}) is the topological charge density operator which is a total divergence. One should also emphasize that the entropy $S$ as well as its surface term $\sim \chi_{E\&M}$ separately  are  gauge invariant observables. Also, while the  term $\sim \chi_{E\&M}$ can be represented as a surface integral, the entropy itself does not  possesses    such  a surface representation. Finally, the entropy (\ref{entropy}) can be interpreted as the entanglement entropy   because the only local observable is $E$, which is constant over space as shown in section \ref{topology}. It means that the measurements of $E$ will be perfectly correlated on the opposite sides of the system\cite{Donnelly:2012st}.

The crucial observation for the present paper is as follows. When the IR physics is properly treated, the entropy (\ref{entropy}) is obviously  a positively defined function. Furthermore, as the   theory  under discussion is ``empty" 
in the sense that it does not describe  any physical propagating degrees of freedom in the bulk, one should expect that the entropy $S$ must vanish in the infinite volume limit $V\rightarrow\infty$.    This expectation follows from the fact that the only dynamics in this system could be  related to the so-called ``edge states" which are localized at the boundary of the system but not in the bulk, similar to other topological field theories \cite{Balachandran:1994vi}. The only way this vanishing result could occur is the presence  of a negative contribution 
which could cancel a conventional positive contribution present in (\ref{entropy}). In different words, the negative contribution in (\ref{entropy})  is a must in order to produce an anticipated  vanishing  entropy in the infinite volume limit $V\rightarrow\infty$.  

As we discussed previously, this contribution with a ``wrong" sign can not be identified with any physical propagating degrees of freedom.
Rather it is related to the tunnelling processes between different topological sectors as discussed in section \ref{topology}.
These discussions again support our claim that     the topological sectors must be included into consideration for self-consistency of the theory. Therefore, the additional terms   they produce leading to the  Topological Casimir Effect  should be considered as   physically observable quantities.  
As the last comment of this Appendix: though  the term (\ref{exact1}) with a ``wrong" sign is a gauge invariant contribution, its  explicit computation    depends on a specific gauge-dependent technique being used. In particular, in the KS framework \cite{KS} this term is saturated exclusively by an unphysical ghost field, 
see explicit computations in \cite{Zhitnitsky:2011tr}. Still, this term (\ref{exact1}) is physical as we argued above, and it can not be   discarded  
on a sole basis that it is saturated by the artificial ghost. 

The main lesson of this Appendix is that    there are extra contributions to the vacuum energy   due to nontrivial topological features of the gauge fields, which do not have counterparts in scalar field theory. Therefore, the standard receipt (that the contribution to the energy and pressure due to the physical Maxwell photons could be obtained directly from expressions for massless scalar field by multiplying the corresponding  scalar expressions by factor two) does not represent a complete description of    the ground state  in the presence of the  gauge fields.

\end{document}